\documentstyle[aps,prl,epsfig,multicol]{revtex}

\begin{document}

\title{ How does the Eurodollar Interest Rate behave?}

\author{ Tiziana Di Matteo and Tomaso Aste$^*$}

\address{INFM-Dipartimento di Fisica, Universit\`a degli Studi di Salerno,}
\address{84081 Baronissi (Salerno), Italy}
\address{e-mail:  tiziana@sa.infn.it .}

\address{ $^*$INFM-Dipartimento di Fisica, Universit\`a di Genova}
\address{via Dodecaneso 33, 16146 Genova, Italy}
\address{AAS, Sal. Spianata Castelletto 16, 16124 Genova, Italy}
\address{e-mail:  aste@fisica.unige.it .}
\date{01/01/01}

\maketitle

\begin{abstract}
An empirical analysis on Eurodollar interest rates daily data in the time period 1990-1996, is performed and compared with Libor data in the time period 1984-1998.
The complementary cumulative distributions for the daily fluctuations at different maturity dates and the Power Spectral Density are computed. 
We find that the probability distribution shows `fat' tails with non-Gaussian behaviours.
Moreover, we study the correlations among Eurodollar interest rates fluctuations with different maturity dates.
By using an original clustering linkage, we show how the collective motion of the interest rates curve can be analyzed in sub-groups of maturity dates with similar behaviours.
\end{abstract}


\begin{multicols}{2}

\section{Introduction } \label{s.1}

The last twenty years have seen a wide development of studies on the interest rates curves. 
So far, most of the efforts in Finance have been devoted to the study of stock prices, where one is dealing with a single object affected by stochastic fluctuations. 
On contrast, the interest rates case deals with several objects that follow similar trends and  can be thought as a curve whose shape is varying in time. 
This makes the subject very challenging since one is no more dealing with the statistics of single objects but with the motion of a whole complex set. 

Nowadays, these studies are becoming very attractive and are approached from many different perspectives \cite{Hull,LibrMant,Wilmott,Rebonato,BouchaudPot}.
Several theoretical models have been proposed  \cite{Vasicek,Cox,Heath,HullWhite,Sornette,Pagan}. 
These models often assume the fluctuations to be Gaussian, neglecting therefore the `fat tail' effects \cite{Embrechts}. 
This is in contrast with the empirical observations \cite{Chan}. 
The inadequacies of the Gaussian models in financial analysis have been reported for a long time \cite{Mandelbrot,MandelbrotB}, but the availability of enormous sets of financial data and the capability to treat them, has recently renewed the interest in the subject  \cite{SornSim,Muller,Mantegna,Ghashghaie,MantegnaSta,Piccinato,Balocchi,Matactz1,Matactz2}. 
While the fat-tail property of the empirical distribution of stocks price changes has been widely documented, the interest rates have been less investigated \cite{Nuyts}. 
Moreover, the studies of interest rates are often limited to only few maturity dates (3 and 6 months cash rates in \cite{MullerDac}, Bund futures in \cite{BouchaudPot}). 
The paper \cite{Bouchaud}, is an exception as the US forward rate curve with maturity dates up to four years is modelled. 

In the present paper, we perform an empirical analysis of time series based on daily prices of Eurodollar future contracts in the time period 1990-1996.  
The data are described in Section \ref{s.2} (more details can be found in the paper of Bouchaud et al. \cite{Bouchaud}). 
Our study is mainly focused on the `tail' regions of the fluctuations probability distributions and on the correlations among time series at different maturity dates.  
The results concerning the probability distribution and the Power Spectral Density of interest rates fluctuations are presented in Section \ref{s.3} and in Section \ref{s.4}, respectively. 
In Section \ref{s.5}, the correlations are analyzed in terms of an original clustering linkage. Finally, some remarks are reported in the last Section.

\section{Empirical data }\label{s.2}
An Eurodollar is a dollar deposited in a bank outside the United States by another bank.
The Eurodollar future contract is a future contract on an interest rate. 
These contracts are traded on an exchange.
The three-month Eurodollar futures traded on the Chicago Merchantile Exchange (CME) is the most popular of the futures contract on short-term interest rates. 
These contracts trade with delivery months of March, June, September, and December up to ten years into the future. 
The variable underlying the contract is the Eurodollar interest rate applicable to a $90$-days period beginning on the third Wednesday of the delivery month.
When the delivery date is reached, and the actual interest rate for the $90$-day period is known, the contract is settled in cash.
The three-month Eurodollar interest rate is the same as the three-month Libor (London Interbank Offered Rate), that is the rate at which large international banks found much of their activities. Specifically, Libor is the rate at which one large bank is willing to lend money to another large bank \cite{Hull,Rebonato,Jamshidian,Zhao}. 
All these rates are determined in trading between banks and change as economic conditions change.

\begin{figure}
\begin{center}
\mbox{\epsfig{file=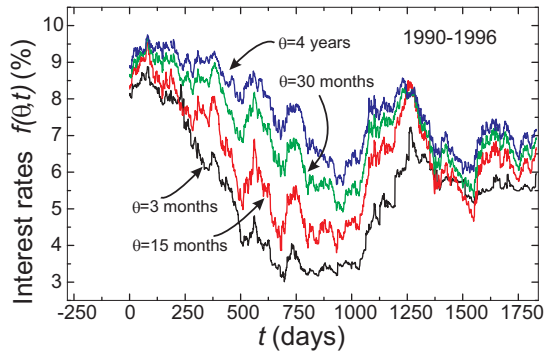,width=2.8in,angle=0}}
\end{center}
\caption{Eurodollar interest rates ($f(\theta,t)$) as function of $t$ for $\theta=3$, 15, 30, 48 months.} 
\label{f.1}
\end{figure}

Following the line of Bouchaud et al. \cite{Bouchaud,CoxIng,Rendleman}, we consider three-months future rates as instantaneous forward rates so obtaining daily values for the forward interest rate curve $f(\theta,t)$, where $t$ is the current date and $\theta$ is the maturity date. 
In the time period analyzed 1990-1996, the variable $t$ counts incrementally the trading days between 1 and $T=$ 1831.
On this time period we have $16$ different time series corresponding to several maturity dates ranging from $\theta=3$ to $48$ months with a step of three months. 
The behaviour of $f(\theta,t)$ as function of $t$ is shown in Fig.\ref{f.1} where, for a better visualisation of the plot, we report only those values corresponding to the following maturity values: $\theta=3,15,30,48$ months. 
The interest rates behaviours for the other maturity dates follow very similar trends in time, and stay mostly inside the shape traced by the two extreme maturity values, namely $\theta=3$ and $\theta=48$ months.

\vspace*{0.5cm}

\begin{figure}
\begin{center}
\mbox{\epsfig{file=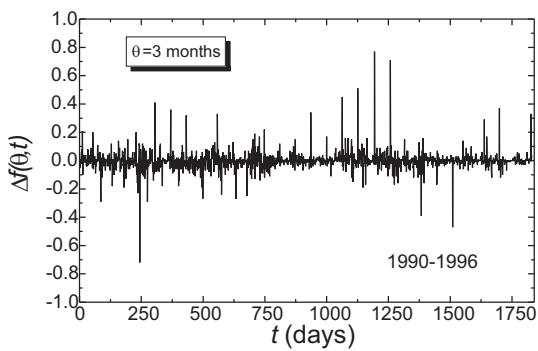,width=2.8in,angle=0}}
\end{center}
\caption{Eurodollar interest rate fluctuation ($\Delta f(\theta,t)$) as function of $t$ for $\theta = 3$ months.} 
\label{f.2}
\end{figure}

The interest rate fluctuations are analyzed by studying the changes from one day to the following day: 
\begin{equation}
\Delta f(\theta,t) = f(\theta, t+\Delta t) - f(\theta,t)\;\;\;\;,
\label{deltaf}
\end {equation}
where $\Delta t=1$ day. 
In Fig.\ref{f.2} these fluctuations are reported as a function of $t$ for $f(\theta=3,t)$. 
All the others at different maturity dates have similar behaviours characterised by stochastic fluctuations around the zero.

\vspace*{0.5cm}

\section{Probability distributions } \label{s.3}

To investigate the nature of the stochastic process underlying the interest rates changes, we first analyze the probability distribution of the daily changes in the interest rates. 

Let us start by the computation of the standard deviation of $\Delta f(\theta,t)$, which is defined as:
\begin {equation}
\sigma (\theta) = \sqrt {{1 \over T_2-T_1} {\sum_{t=T_1}^{T_2}} (\Delta f(\theta,t)- < \Delta f >)^2}  \;\;\;,
\end {equation}
where $T_1$ and $T_2$ delimit the range of $t$, and $<\Delta f >$ is the average over time of $\Delta f(\theta,t)$ (which tends to zero for $T_2-T_1 \rightarrow \infty$).
In the whole period 1990-1996, $\sigma (\theta)$ v.s. $\theta$ has a minimum value of $0.063$ at $\theta=3$ months, then increases up to a maximum value of $0.087$ at $\theta=$ 1 year, once reached this maximum the value slowly decreases until $\sigma (\theta) = 0.062$ at $\theta = 4$ years. 

\vspace*{0.5cm}
\begin{figure}
\begin{center}
\mbox{\epsfig{file=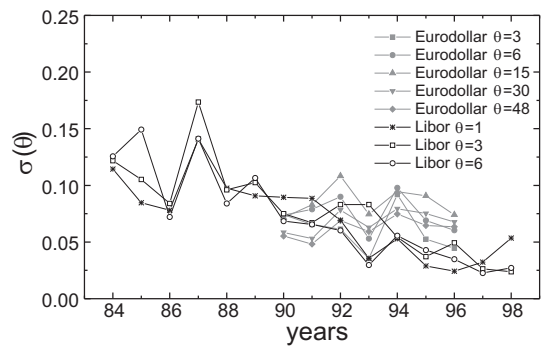,width=2.8in,angle=0}}
\end{center}
\caption{ Standard deviation $\sigma(\theta)$ as function of the years, for Eurodollar at $\theta= 3$, 6, 15, 30, 48 months (time period 1990-1996) and Libor for $\theta=1$, 3, 6 months (time period 1984-1998).} 
\label{f.3}
\end{figure}

We compare the Eurodollar standard deviation of $\Delta f$ for each single year between 1990-1996 with the one of the Libor data corresponding to a longer time period from 1984 to 1998.  
In Fig.\ref{f.3} the values of $\sigma (\theta)$ in the different years are plotted for several maturity dates. 
The standard deviations for both Eurodollar and Libor series with the same maturity dates $\theta = $3 and 6 months, have similar values and show  decreasing trends between 1984 to 1998. 

The statistical distributions of the daily changes in the interest rates are better analyzed by computing their complementary cumulative distributions $\Psi(\Delta f)$, that tell us the probability to find a daily change which is \emph{larger than} $\Delta f$:
\begin {equation}
\Psi(\Delta f) = 1 - {\int^{\Delta f}_{-\infty}} { p(\xi) d \xi}  \;\;\;;
\end {equation}
with $p(\xi)$ being the probability density distributions of $\Delta f(\theta,t)$.

\vspace*{0.5cm}
\begin{figure}
\begin{center}
\mbox{\epsfig{file=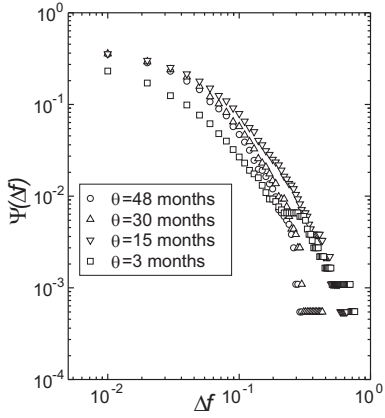,width=2.in,angle=0}}
\end{center}
\caption{Complementary cumulative distribution $\Psi (\Delta f)$ v.s. $\Delta f$ for Eurodollar data at different maturity dates: $\theta= 3$, 15, 30, 48 months.} 
\label{f.4}
\end{figure}

We find that our time series are highly leptokurtic for all the maturity dates and are characterized by non-Gaussian profiles for large interest rates changes.
The results for $\Psi(\Delta f)$ are reported in Fig.\ref{f.4} for the tails regions for positive changes of $\Delta f$ with $\theta=3,15,30,48$ months.
These curves show, in a log-log scale, three distinct parts characterised by different decreasing trends: i) a first region ($0.01<|\Delta f |< 0.05$) with slopes between $-0.4 \div -0.6$; ii) an intermediate region ($0.05 < |\Delta f | < 0.2$) with a faster decreasing behaviour (slopes $-1.6 \div -2.7$); iii) a final region ($| \Delta f | > 0.2$) which is characterized by larger slopes (between -2.3 up to -10).
Comparable results are obtained for all the time series at different maturity dates.

\vspace*{0.5cm}
\begin{figure}
\begin{center}
\mbox{\epsfig{file=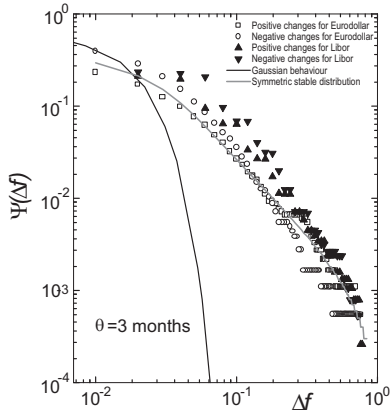,width=2.in,angle=0}}
\end{center}
\caption{ Complementary cumulative distribution ($\Psi (\Delta f)$) and cumulative distribution (1-$\Psi (\Delta f)$) v.s. $|\Delta f |$, for Eurodollar and Libor data at $\theta=3$ months.
The grey-line refers to the Gaussian behaviour, whereas the black-line is the L\'evy stable distribution.} 
\label{f.5}
\end{figure}

In Fig.\ref{f.5}, we plot the tail region of $\Psi (\Delta f)$ at $\theta=3$ months as function of both positive and negative changes of $\Delta f$. 
We observe that the probability distributions are slightly asymmetric with negative changes that are slightly more probable than the positive ones. 
On the other hand, the large variations result more probable for the positive changes, and the overall distribution leads -correctly- to a zero mean value for $\Delta f$. 
In the same figure the two empirical curves for the Eurodollars are compared with the two curves for the Libor at $\theta=3$ months. 
Large fluctuations in Eurodollar and Libor data show very similar behaviours with comparable probability values. 
In these distributions the tails are non-Gaussian: they are fatter than the Gaussian ones. 

Instead that with the Gaussian distribution, a more appropriate comparison should be performed with the general class of L\'evy, Khinthine stable distributions \cite{Gennady}. 
A symmetric stable distribution with zero mean can be written as \cite{LibrMant}: 
\begin {equation}
P_{L} (x) \equiv {1 \over \pi} \int_{0}^{\infty} e^{-\gamma q^{\alpha}} \cos(qx) dq      \;\;\;,
\label{Levy}
\end {equation}
where $0< \alpha \le 2$ and  $\gamma$ is a positive scale factor. 
For $\alpha = 2$ and  $\gamma = \sigma^2/2$, $P_L$ is a Gaussian distribution with zero mean and standard deviation equal to $\sigma$; whereas for $\alpha = 1$, $P_L$ is the Lorentzian distribution. 

In Fig.\ref{f.5} we compare the Eurodollar empirical data with a symmetric stable distribution, given by Eq.(\ref{Levy}), having $\alpha = 1.45$. 
One can note that these curves have a tail region that seems to follow a L\'evy stable distribution. But we may expect a faster decrease to zero for larger fluctuations of $\Delta f$, as for instance the one described by the truncated L\'evy distribution \cite{MS}. 
A deeper analysis on this aspect will be done in a future work.
Let note that all time series at different maturity dates show behaviours analogous to the one at $\theta=3$ months.

\vspace*{0.5cm}
\begin{figure}
\begin{center}
\mbox{\epsfig{file=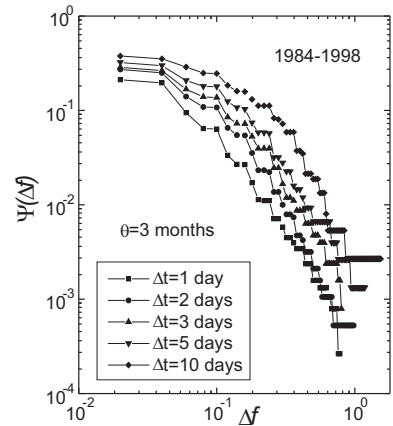,width=2.in,angle=0}}
\end{center}
\caption{Complementary cumulative distribution $\Psi (\Delta f)$ v.s. $\Delta f$ for Libor data at the maturity date $\theta= 3$ months, for different values of the time increment $\Delta t$.} 
\label{f.6}
\end{figure}

Fig.\ref{f.6} reports $\Psi (\Delta f)$ for the Libor non-overlapping $\Delta f$ time series with $\theta = 3$ month and $\Delta t$ values ranging between $1$ and $10$ (see Eq.(\ref{deltaf})). 
Note that the Gaussian behaviour tends to appear in the tail regions when the $\Delta t $ increases.

\section{Spectrum analysis } \label{s.4}

\begin{figure}
\begin{center}
\mbox{\epsfig{file=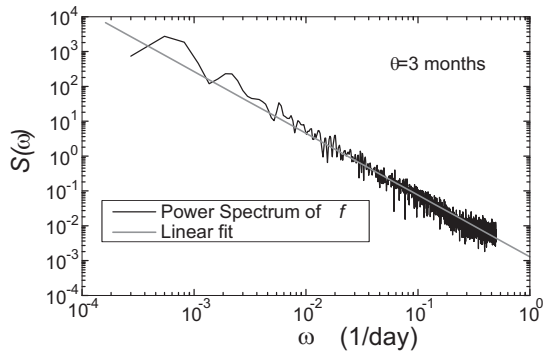,width=2.8in,angle=0}}
\end{center}
\caption{Power spectrum $S(\omega)$ of the Eurodollar interest rates at $\theta = 3$ months, as a function of the frequency $\omega$.} 
\label{f.7}
\end{figure}

In this section we investigate the statistical properties of our time series in the frequency domain. To this end we compute the power spectral density (PSD) \cite{Kay} of our random variables 
by using the periodogram approach to PSD estimation, that is currently one of the most popular and computationally efficient PSD estimator. 
This periodogram spectral estimate is obtained as the squared magnitude of the output values from an FFT performed directly on the data set. 
The power spectrum analysis for $f(\theta,t)$ indicates a stocastic signal with spectral components which are decreasing as a power law:
\begin {equation}
S(\omega) \sim {\omega}^{-\beta} \;\;\;,
\end {equation}
with $\beta \cong 1.8$. 
Fig.\ref{f.7} shows the power spectrum of the spot interest rate $r(t)=f(\theta=3,t)$ on a log-log scale. 
The slope of the linear fit is $\beta = -1.77 \pm 0.01$ and, within the error, it is the same for all the other Eurodollar interest rates at different maturity values, and it is in good agreement with the Libor at $\theta = 1$, $3$, and $6$ months which have $\beta = -1.80 \pm 0.01$. 

\vspace*{0.5cm}

\begin{figure}
\begin{center}
\mbox{\epsfig{file=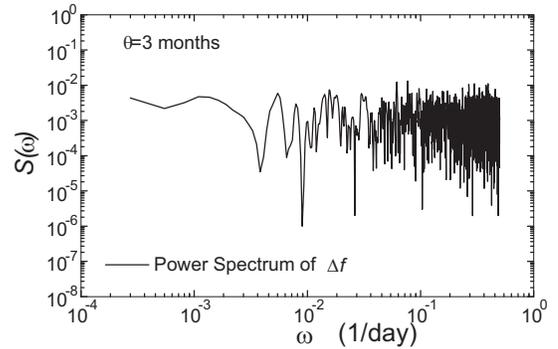,width=2.8in,angle=0}}
\end{center}
\caption{ Power spectrum $S(\omega)$ of the Eurodollar interest rates fluctuations  at $\theta = 3$ months, as a function of the frequency $\omega$.} 
\label{f.8}
\end{figure}

The same analysis for $\Delta f(\theta,t)$ (see Fig.\ref{f.8}) shows a flat spectrum, 
from which we can infer that the interest rates changes are ``pair wise uncorrelated'' and the stochastic process is approximately a white noise.

\vspace*{0.5cm}

\section{Correlations } \label{s.5}

In order to investigate the correlations among the interest rates daily changes at different maturity dates, let us here define the cross-correlation matrix:
\begin {equation}
c_{i,j} = {{ <\Delta f_i \Delta f_j> - <\Delta f_i> <\Delta f_j>} \over {\sigma_i \sigma_j} } \;\;\;,
\label{cij}
\end {equation}
where $i,j=1,...,16$, $\Delta f_i$ indicates the interest rate daily changes for maturity $\theta_i=3 i$ months, and $\sigma_i$ is the standard deviation of interest rates changes $\Delta f_i$. 
The symbol $<...>$ denotes a time average performed over all the days of the investigated period. 
The correlation coefficients are computed between all the pairs of indices labelling our Eurodollar series. 
Therefore we have a $16 \times 16$ symmetric matrix with $c_{i,i}=1$ on the diagonal. 

The daily changes in the interest rates are highly correlated with coefficients in the range $0.48 \leq c_{i,j} \leq 0.98$ (for $i \neq j$). 
These strong correlations between the interest rates fluctuations suggest that one can consider their behaviour as a collective motion of a set of quantities joined together with some interaction force. 

\vspace*{0.5cm}
\begin{figure}
\begin{center}
\mbox{\epsfig{file=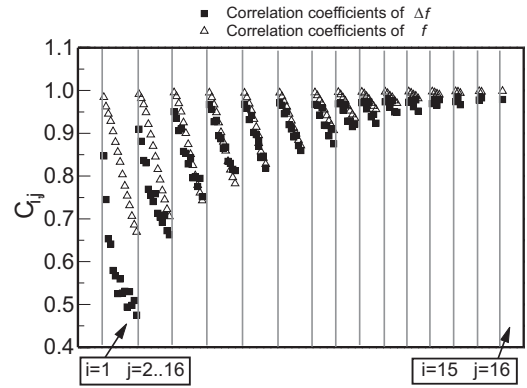,width=2.8in,angle=0}}
\end{center}
\caption{ Correlation coefficients ($c_{i,j}$, with $i,j =1..16$ and $i<j$) for $\Delta f(\theta,t)$ and $f(\theta,t)$. } 
\label{f.9}
\end{figure}

The correlation coefficients $c_{i,j}$ for $\Delta f_i = \Delta f(\theta_i,t)$ are shown in Fig.\ref{f.9} together with the correlation coefficients for $f(\theta_i,t)$ (which are obtained substituting $\Delta f(\theta_i,t)$ with $f(\theta_i,t)$ in Eq.(\ref{cij})).
The plot reads in the following way: the vertical lines group together the correlation coefficients $c_{i,j}$ for a fixed value of $i$ and $j$ ranging between $i+1$ to $16$. 
Looking at the plot from left to right, $i$ increases its value from $i=1$ up to $i=15$. 
The coefficients $c_{i,j}$ have decreasing trends with $j$. 
This corresponds to the intuitive fact that interest rates with close maturity dates are more correlated than interest rates with distant maturity (e.g. $c_{1,2}> c_{1,4}> c_{1,8}> c_{1,16}$). 
Such decreasing trends support the idea that interest rates are directly interacting between neighbours only (which means that the maturity $\theta_i$ has direct interactions with maturity $\theta_{i \pm 1}$), so making a sort of `string'  \cite{Sornette,Bouchaud,Gopi}. 

A closer look at Fig.\ref{f.9} shows that the decreasing trend in the correlation coefficients for $\Delta f_i$, is not perfectly monotonic. 
In particular, one can observe deviations from the monotonic trend for correlation coefficients involving $j \gtrsim 8$ (i.e. maturity dates $\theta_j \gtrsim 2$ years). 

To understand the geometrical and topological structure of the correlation coefficients, we introduce the metric distance $d_{i,j}$ \cite{RMant} between the series $\Delta f_i $ and $\Delta f_j$ which is defined as:
\begin {equation}
d_{i,j} = {\sqrt{2(1-c_{i,j}) }}   \;\;\;.
\end {equation}
By definition, $c_{i,j}$  is equal to zero if the interest rates series $i$ and $j$ are totally uncorrelated, whereas $c_{i,j} = \pm 1$ in the case of perfect correlation/anti-correlation. 
Therefore,
\begin {eqnarray}
d_{i,j} &= 0  \;\;\;\;\;\; &if  \;\;\;   c_{i,j} = 1 \nonumber \\
d_{i,j} &= \sqrt 2 \;\;\; &if \;\;\;    c_{i,j} = 0 \\
d_{i,j} &= 2 \;\;\;\;\; &if \;\;\;        c_{i,j} = -1   \nonumber \;\;\;.
\end {eqnarray}
Note that this metric distance fulfils the three axioms of a metric: 1) $d_{i,j}=0$ if and only if $i=j$; 2) $d_{i,j}= d_{j,i}$; 3) $d_{i,j} \leq d_{i,k}+d_{k,j}$. Indeed, $d_{i,j}$ is the Euclidean distance, in a $T$-dimensional space, between the two vectors with components $ [ \Delta f (\theta_{i},1),$$ \Delta f (\theta_{i},2),$$.....,$ $\Delta f (\theta_{i},T)] / \sigma_{i}$
and
$[ \Delta f (\theta_j,1),$$.... ,$$\Delta f (\theta_j,T)] / \sigma_{j}$.
These vectors have unitary lengths. Therefore, they connect the centre of a unit sphere, in a $T$-dimensional space, with points on its surface. 
Similar behaviours of the interest rate changes correspond to two close points on the sphere ($d_{i,j} \ll 1$). 
Uncorrelated behaviours correspond to two vectors pointing toward perpendicular directions ($d_{i,j} = \sqrt 2$). 
Whereas anti-correlated behaviours are associated to two vectors pointing toward two opposite directions ($d_{i,j} = 2$= {\it sphere diameter}). 

We study the geometrical and topological arrangements of Eurodollar interest rates by introducing the ultra-metric distance $\hat d_{i,j}$ which satisfies the first two properties of the metric distance and replaces the triangular inequality with the stronger condition: $\hat d_{i,j} \leq max [ \hat d_{i,k},\hat d_{k,j}]$, called `ultra-metric inequality'. 
Once the metric distance $d_{i,j}$ between our interest rates fluctuations is known, one can introduce several ultra-metric distances. 

Mantegna et. al have used the `subdominant ultra-metric', obtained by calculating the minimum spanning tree  \cite{MST} connecting the stock indices \cite{RMant,Bonanno}. 
In this paper we consider a different ultra-metric space that emphasizes the cluster-structure of the data. 

In our case, a ``cluster'' is a set of elements with relative distances $d_{i,j}$ which are smaller than a given threshold distance $\bar \delta$, whereas disjoined clusters have some elements which are at distances larger than $\bar \delta$.  
Let us define the \emph{size} of a cluster as the number of elements that it contains, and its \emph{diameter} as the maximum relative distance between them. 
Therefore, an isolated element can be considered a cluster of size 1 and zero diameter.

We define the \emph{ultra-metric distance} $\hat d_{i,j}$ between two elements $i,j$ belonging to two different clusters as the maximum metric distance between all the couples of elements in the two clusters. Therefore, the ultra-metric distance between two isolated elements (clusters of size $1$) coincides with the metric distance. 
Whereas, the ultra-metric distance between two elements inside the same cluster is equal to the cluster diameter. 

The linkage procedure that we adopt is described in details in Appendix \ref{A.1}. 
This procedure yields to a set of clusters which have diameters which are smaller or equal than the threshold distance $\bar \delta$, and have relative ultra-metric distances $\hat d$ which are larger than $\bar \delta$. 

\vspace*{0.5cm}
\begin{figure}
\begin{center}
\mbox{\epsfig{file=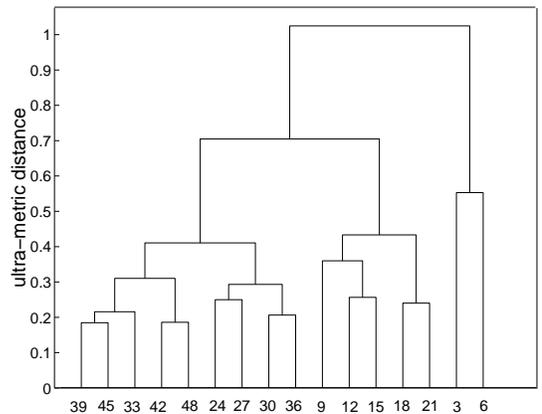,width=2.8in,angle=0}}
\end{center}
\caption{Hierarchical tree obtained from the correlation coefficients of the 16 Eurodollar interest rates fluctuations time series in the time period 1990-1996. (On the $x$-axis are reported the maturity dates and on the $y$-axis the ultra-metric distances.)} 
\label{f.10}
\end{figure}

The result of this recursive linkage procedure, for the $16$ maturity dates ($\theta=3,..,48$) in the whole time period $1990$-$1996$, is reported in the hierarchical tree shown in Fig.\ref{f.10}.
The clustering begins with maturities $39$-$45$ that gather together at $\bar \delta=0.185$; at a little larger value $\bar \delta=0.186$, the couple $42$-$48$ makes another cluster and the couple $30$-$36$ merges at the $\bar \delta$= 0.206; then the maturity $33$ joins the first cluster at a $\bar \delta= 0.216$. 
The clustering process goes on (see Fig.\ref{f.10}) and ends when only one large cluster is formed at $\bar \delta = 1.02$. 

\vspace*{0.5cm}
\begin{figure}
\begin{center}
\mbox{\epsfig{file=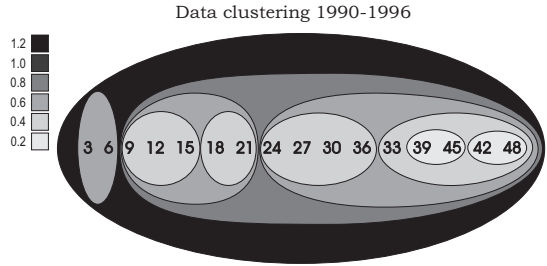,width=2.8in,angle=0}}
\end{center}
\caption{ Grey-scale-tones visualisation of the clusters at different $\bar \delta$ values, for the 16 Eurodollar interest rates fluctuations time series in the time period 1990-1996.} 
\label{f.11}
\end{figure}

This mechanism of nucleation, growth, and coalescence is better visualised in Fig.\ref{f.11} where the clusters are explicitly drawn with different grey-tones for $6$ distinct ranges of $\bar \delta$ ($0 \div 0.2$; $0.2 \div 0.4$; $0.4 \div 0.6$; $0.6 \div 0.8$; $0.8 \div 1$; $1 \div 1.2$). 
It is evident from Figs.\ref{f.10} and \ref{f.11} that, for ultra-metric distances $\bar \delta$ in the range between $0.4$ and $0.6$, the data set is gathered into $3$ main clusters: $Cls_1 = \{ 3,6 \}$, $ Cls_2=\{9,12,15,18,21\}$, $ Cls_3=\{24,27,....,45,48\}$, with diameters $\hat d =$ 0.553, 0.411, 0.433, respectively. 
The first cluster ($Cls_1$) gathers together $\Delta f(\theta,t)$ with maturity shorter than $1$ year; $Cls_2$ contains those with maturity dates between $1$ year and $2$ years; whereas $Cls_3$ has those with maturity dates which are larger than $2$ years. 

\vspace*{0.5cm}
\begin{figure}
\begin{center}
\mbox{\epsfig{file=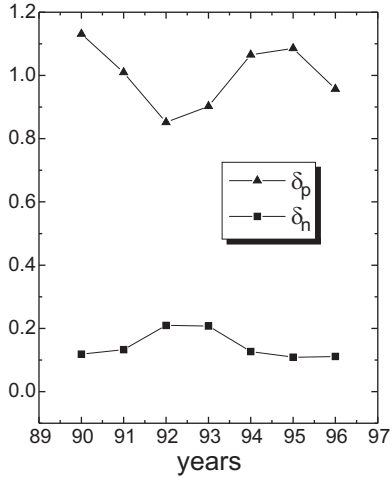,width=2.in,angle=0}}
\end{center}
\caption{ The nucleation distance $\delta_n$ and the distance $\delta_p$ as functions of the years between 1990-1996.} 
\label{f.12}
\end{figure}

Let us now proceed looking at the clusters arrangement of $\Delta f(\theta,t)$ for each single year between $1990$ to $1996$.
We find that the cluster arrangement in each single year is similar to the one observed in Figs.\ref{f.10}, \ref{f.11} where the whole time period is considered. 
First we have the nucleation of a couple of maturity dates at distances $\bar \delta = \delta_n$ whose values are between $0.109$ and $0.210$ (see Fig.\ref{f.12}, where these nucleation distances are plotted for each year). 
Then we observe the growth of several clusters, that in the distances range between $0.3$ to $0.6$, makes a structure of four well identifiable clusters. 
The first one is composed by the maturity $\theta=3$ and sometime $\theta=6$. 
The second includes time series with maturity date values around $1$ year ($\theta=6 \div 15$). 
The third gathers maturity dates between 2 and 2.5 years ($\theta=12 \div 33$) and the fourth the largest maturity dates $> 2$ years ($\theta=27 \div 48$). 
Finally, these clusters coalesce into a single one at distances $\bar \delta = \delta_p$, whose values (ranging between $0.852$ and $1.131$) are plotted in Fig.\ref{f.12} for each year.
Note that the values of $\delta_n$ and $\delta_p$, calculated for the whole time period (1990-1996), are within the range of the values plotted in Fig.\ref{f.12}.

\section{Conclusions}

We have empirically analyzed Eurodollar interest rates at different maturity dates in the time period 1990-1996.
We observed that the tail region has an evident non-Gaussian behaviour - a crucial feature for the risk management analysis. 
Therefore, by using a Gaussian probability density function one underestimates the probability of large interest rates fluctuations. 
These results are confirmed by the same analysis done on Libor time series in the longer time period from $1984$ to $1999$. 
Moreover, a power spectrum analysis has been carried out in order to give some insights into the stochastic process underlying the examined interest rates time series.
The power spectrum for the interest rates fluctuations shows a white-noise-like flat behaviour.
Whereas, the one for the interest rates has a power law spectrum with exponent $1.77$ (Eurodollar), and $1.80$ (Libor) for all the different maturity dates.

We have also analyzed the correlations between interest rates fluctuations at different maturity dates by using a linkage procedure that emphasizes the cluster structure. 
In the time period 1990-1996, we have found the formation of three clusters joining together time series corresponding to three maturity dates periods: shorter than 1 year, between 1 to 2 years, larger than 2 years. 
A cluster structure with similar features have been found for each single year.
The high correlation values that we have found, confirm that the interest rates fluctuations behaviour at different maturity dates must be seen as a collective motion.
On the other hand, the result from the clustering linkage procedure indicates that the deviations from this collective behaviour can be gathered in some maturity-groups.
This might suggest that theoretical models for the interest rates behaviours should usefully consider different kinds of interactions in different sub-groups of maturity dates.

\bigskip
{\Large \bf Acknowledgements}

\noindent
Many thanks to Jean-Philippe Bouchaud, Michel Dacorogna, Rosario Mantegna,  for providing us the data, and for many helpful discussions, comments and indications.
T. Di Matteo wishes to thank Sandro Pace for fruitful discussions and support.
T. A. acknowledges the logistic support from AAS (Genova).

\appendix

\section{Linkage procedure } \label{A.1}

At the starting point of our linkage procedure we have a set of $N$($=16$) isolated elements which can be considered as $N$ clusters of size equal to $1$ and diameter equal to zero.
We define the ultra-metric distance $\hat d_{i,j}$ between two elements $i,j$ belonging to two different clusters as the maximum metric distance between all the couples of elements in two clusters. Whereas, the ultra-metric distance between two elements inside the same cluster is equal to the cluster diameter. 

The first step of the linkage procedure finds the couple of elements $i,j$ which are at the smallest distance $\hat d_{i,j}$. 
This couple is linked together in a new cluster of size $2$ and diameter $\hat d_{i,j}$. 
Then we seek for the pair of elements with the next-smallest ultra-metric distance, and we generate a new cluster, or we enlarge the previous one if one of the two clusters belongs to that cluster. 
This procedure is recursively iterated for increasingly larger distances by joining isolated elements to clusters or by merging together larger clusters.
The procedure terminates when all the elements are jointed together in a sole cluster. 
This ending diameter coincides with the largest metric distance between two elements in the data set.

\end{multicols}

\end{document}